\documentclass[aps,prl,twocolumn,amsmath,amssymb,superscriptaddress]{revtex4}
\usepackage{graphicx}
\usepackage{amssymb}
\usepackage{natbib}
\usepackage[compact]{titlesec}
\usepackage[rawfloats=true]{floatrow} 
\usepackage{array}
\usepackage{chngcntr}
\restylefloat{figure}

\begin{document}

\title{Neutron spin resonance as a probe of Fermi surface nesting and superconducting gap symmetry in Ba$_{0.67}$K$_{0.33}$(Fe$_{1-x}$Co$_{x}$)$_{2}$As$_{2}$}
\author{Rui Zhang}
\affiliation{Department of Physics and Astronomy, Rice University, Houston, Texas 77005, USA}
\author{Weiyi Wang}
\affiliation{Department of Physics and Astronomy, Rice University, Houston, Texas 77005, USA}
\author{Thomas A. Maier}
\affiliation{Computer Science and Mathematics Division and Center for Nanophase Materials Sciences,Oak Ridge National Laboratory, Oak Ridge, Tennessee 37831-6494, USA}
\author{Meng Wang}
\affiliation{School of Physics, Sun Yat-Sen University, Guangzhou 510275, China}
\author{Matthew B. Stone}
\affiliation{Neutron Scattering Division, Oak Ridge National Laboratory, Oak Ridge, Tennessee 37831, USA}
\author{Songxue Chi}
\affiliation{Neutron Scattering Division, Oak Ridge National Laboratory, Oak Ridge, Tennessee 37831, USA}
\author{Barry Winn}
\affiliation{Neutron Scattering Division, Oak Ridge National Laboratory, Oak Ridge, Tennessee 37831, USA}
\author{Pengcheng Dai}
\email{pdai@rice.edu} 
\affiliation{Department of Physics and Astronomy, Rice University, Houston, Texas 77005, USA}

\begin{abstract}
{We use inelastic neutron scattering to study energy and wave vector dependence of the superconductivity-induced resonance
 in hole-doped Ba$_{0.67}$K$_{0.33}$(Fe$_{1-x}$Co$_{x}$)$_{2}$As$_{2}$ ($x=0,0.08$ with $T_c\approx 37, 28$ K, respectively). In previous work on electron-doped Ba(Fe$_{0.963}$Ni$_{0.037}$)$_2$As$_2$ ($T_N=26$ K and $T_c=17$ K),
the resonance is found to peak sharply at the antiferromagnetic (AF) 
ordering wave vector ${\bf Q}_{\rm AF}$ along the longitudinal direction, 
but disperses upwards away from ${\bf Q}_{\rm AF}$ along the transverse direction [Kim {\it et al.}, Phys. Rev. Lett. {\bf 110}, 177002 (2013)].  
For hole doped $x=0, 0.08$ without AF order, we find that 
the resonance displays ring-like upward dispersion away from
${\bf Q}_{\rm AF}$ along both the longitudinal and transverse directions.  
By comparing these results with calculations using the random phase approximation,
we conclude that the dispersive resonance is a 
direct signature of isotropic superconducting gaps arising from  
nested hole-electron Fermi surfaces. }
\end{abstract}

\maketitle

Understanding the interaction between magnetism and unconventional superconductivity continues to be an important topic in 
modern condensed matter physics \cite{scalapino,BKeimer,dai,thompson}. In copper and iron-based high-transition-temperature (high-$T_c$)  superconductors, the parent compounds are long-range ordered antiferromagnets and superconductivity arises from 
electron or hole-doping to the parent compounds \cite{scalapino,BKeimer,dai}. Although static 
antiferromagnetic (AF) order in the parent compounds is gradually suppressed with increasing doping, dynamic spin correlations (excitations) remain and inelastic neutron scattering (INS) experiments have identified a ubiquitous collective spin excitation mode, termed neutron spin resonance, 
that occurs below $T_c$ with a temperature-dependence similar to the superconducting order parameter \cite{J.Rossat-Mignod,eschrig,christianson,lumsden,schi09,dsinosov09,C.Zhang,stock08}. 
Moreover, the energy of the resonance has been associated with $T_c$ or superconducting gap size $\Delta$ \cite{wilson,D.Inosov,G.Yu,QSWang16}, thus establishing its direct connection with 
superconductivity. 
For hole-doped copper oxide superconductors such as YBa$_2$Cu$_3$O$_{6+x}$, the resonance, obtained by subtracting the normal-state spin 
excitations from those in the superconducting state, displays predominantly a downward dispersion 
away from the in-plane AF ordering wave vector ${\bf Q}_{\rm AF}=(1/2,1/2)$ of the proximate 
tetragonal phase \cite{bourges00,dai00,Reznik04,Hayden04}.
In the case of undoped iron pnictides, the AF order occurs in the orthorhombic lattice with spins aligned anti-parallel along
the orthorhombic $a_o$ axis ($H$ direction in reciprocal space) and parallel along the $b_o$ axis ($K$ direction) at the in-plane wave vector at ${\bf Q}_{\rm AF}=(1,0)$ \cite{dai}. 
Here, the resonance for electron-underdoped iron pnictide BaFe$_{1.926}$Ni$_{0.074}$As$_2$ with coexisting AF order and
superconductivity \cite{xylu13} is centered around ${\bf Q}_{\rm AF}$ along the 
$H$ (longitudinal) direction but has an upward
spin-wave-like dispersion along the $K$ (transverse) direction \cite{Kim13}. Finally, for heavy Fermion superconductor CeCoIn$_5$ \cite{thompson}, the resonance exhibits a spin-wave ring like upward dispersion \cite{raymond15,song16} reminiscent of spin waves 
in nonsuperconducting CeRhIn$_5$ \cite{Das14,stock15}.

\begin{figure}[t]
\includegraphics[scale=.4]{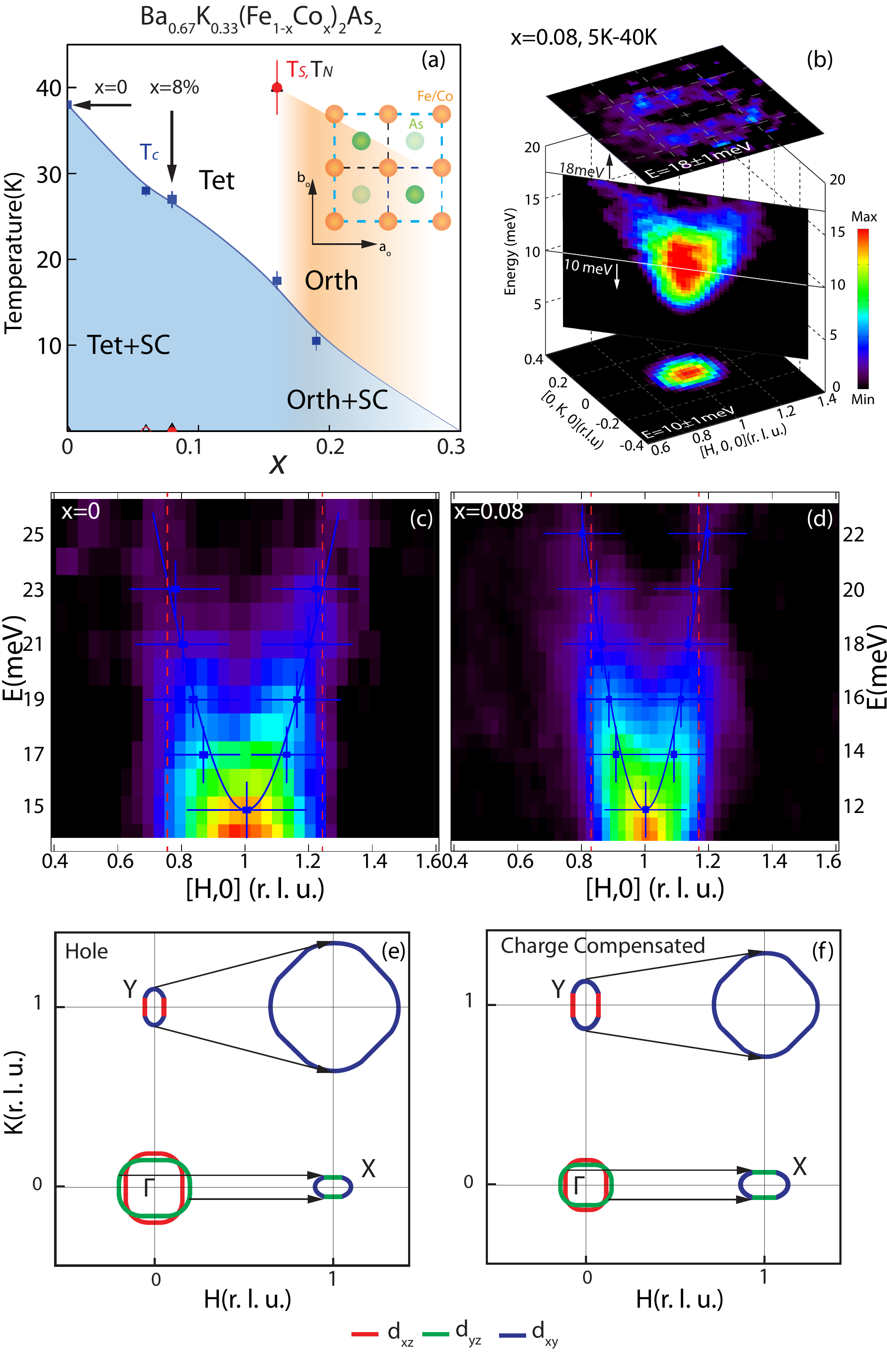}
\caption{
(a) Electronic phase diagram of Ba$_{0.67}$K$_{0.33}$(Fe$_{1-x}$Co$_{x}$)$_{2}$As$_{2}$, black arrows indicate the Co-doping samples reported in this work. Blue squares, red triangles, black squares represent $T_c$, $T_s$, $T_N$ respectively.
(b) The 3D plot of the resonance dispersion of the $x=0.08$ sample in reciprocal space 
after correcting the Bose population factor. 
The orange area marks possible long range AF ordered phase induced by Co-doping.
The bottom and top slice shown in this plot are energy integrated at $E=10\pm 1$  and $18\pm 1$ meV with $E_i$=35 meV, respectively.  
(c,d) Constant wave vector slice of the resonance from 10 meV to 22 meV with $E_i=35$ meV 
along the $H$ direction for $x=0$ and 0.08, respectively. The slice is integrated from $-0.15\leq K \leq 0.15$. The blue solid lines are fits of resonance dispersion 
using equation (1). The red dashed line indicates the width of the resonance dispersion at 8 meV above its initial energy. All scattering intensities in Figs. 1-4 are corrected by the magnetic
form factor and Bose population factor. (e,f) Co-doping evolution of the Fermi surfaces from density functional theory calculation, where red, green, and blue indicate
$d_{xz}$, $d_{yz}$, and $d_{xy}$ orbitals. 
}
\end{figure}

Although it is generally accepted that the presence of a resonance is a signature of unconventional superconductors \cite{scalapino},
there is no consensus on its microscopic origin. The most common interpretation of the resonance is that it is 
 a spin-exciton, arising from particle-hole excitations involving momentum states near the Fermi surfaces that possess
opposite signs of the $d$-wave \cite{eschrig,Abanov99,chubukov01} or $s^{\pm}$-wave \cite{hirschfeld} superconducting order parameter. For 
$d_{x^2-y^2}$-wave superconductors such as copper oxides \cite{scalapino} and CeCoIn$_5$ \cite{thompson}, the resonance is expected to
show a downward dispersion away from ${\bf Q}_{\rm AF}=(1/2,1/2)$ of their parent compounds \cite{eschrig}. Therefore, the surprising observation of
a spin-wave ring like upward dispersion of the resonance in CeCoIn$_5$ \cite{song16} suggests that the mode is a magnon-like excitation
revealed in the superconducting state due to reduced hybridization between $f$ electrons and conduction electrons,
and not an indication of a sign reversed order parameter \cite{chubukov08}.  For iron pnictide superconductors [Fig. 1(a)], the resonance is generally believed 
to be a spin exciton arising from sign-reversed quasiparticle excitations between the hole and electron Fermi surfaces located at $\Gamma$ and $X/Y$ points in reciprocal space, respectively [Fig. 1(e)] 
\cite{hirschfeld}. Although the observation of a transverse upward spin-wave-like dispersive resonance in superconducting BaFe$_{1.926}$Ni$_{0.074}$As$_2$ is different from the downward dispersion of the mode
in YBa$_2$Cu$_3$O$_{6+x}$, it has been argued that the mode is a spin exciton arising 
from isotropic $s^{\pm}$ superconducting gaps at $\Gamma$ and $X/Y$ points and its coupling with the normal state
spin fluctuations via 
\begin{equation}
\Omega_\textbf{q}=\sqrt{\Omega_0^2+c_{res,\textbf{q}}^2{\bf q}^2}
\end{equation}, where $\Omega_0$ is the
resonance energy, $c_{res,\textbf{q}}=\Omega_0\xi_\textbf{q}$ is the velocity of the resonance and its anisotropy in momentum ($\textbf{q}$) space is due to the anisotropy in the normal state spin-spin correlation length $\xi_\textbf{q}$ \cite{Kim13,tucker12,Luo12}. However, 
spin waves from static AF order coexisting  with superconductivity in BaFe$_{1.926}$Ni$_{0.074}$As$_2$ may complicate such interpretation.

\begin{figure}[t]
\includegraphics[scale=.50]{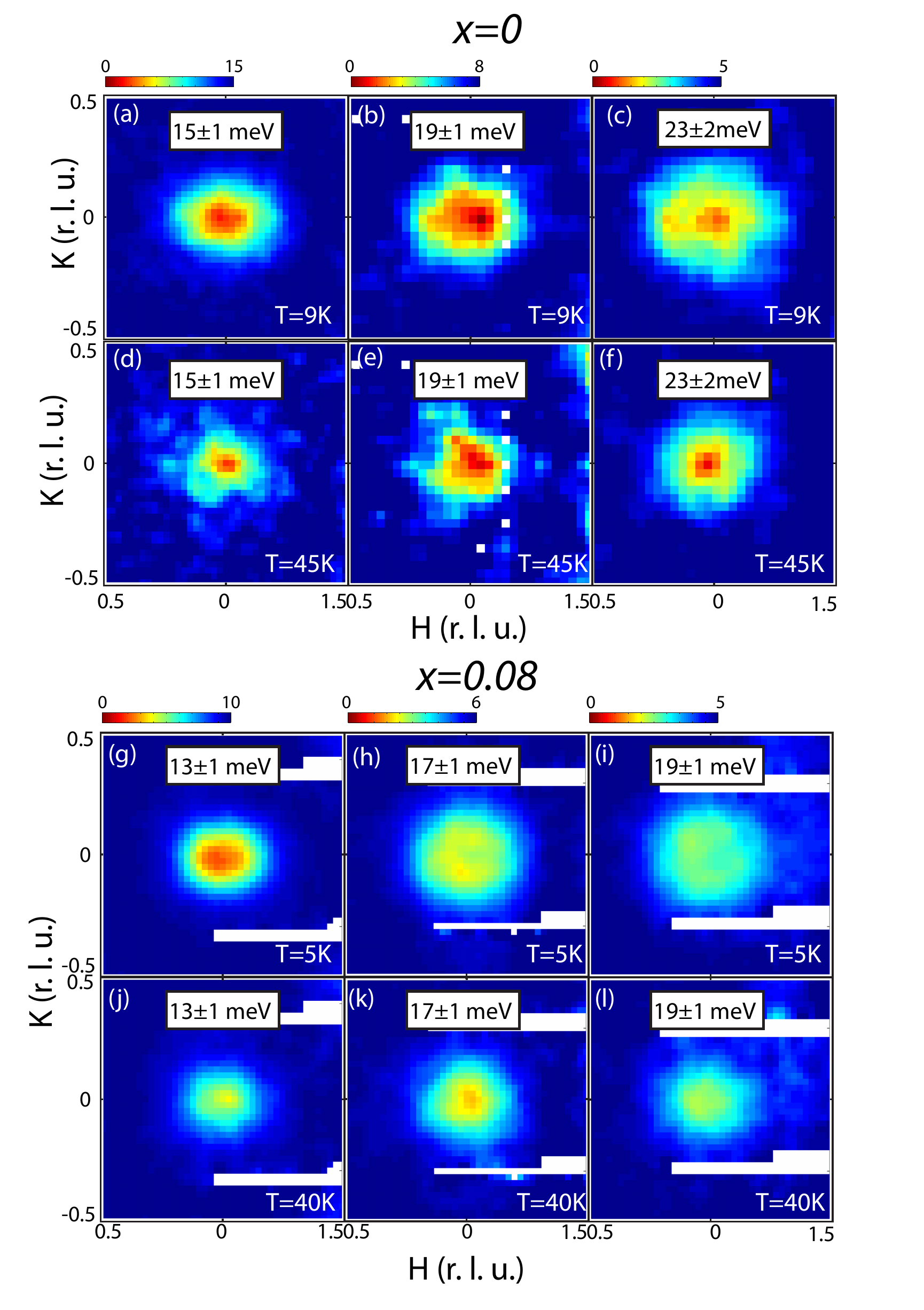}
\caption{The 2D images of spin excitations as a function of energy below and above
$T_c$ for $x=0$ (a-f) and $0.08$ (g-l). Cuts (a), (b), (d), (e), (g)~(l) are taken with $E_i=35$ meV and (c) and (f) are taken with $E_i=70$ meV. The instrument energy resolutions for $E_i=35$ meV 
and $E_i=70$ meV are $\sim$1.5 and 3.5 meV, respectively. The white areas in Figs. 2-4 are due to
dead detectors. 
}
\end{figure}

If the resonance in iron pnictide superconductors is indeed a spin exciton without
related to spin waves from static AF order, one would expect that modifying the wave vector dependence of the normal state spin fluctuations should affect the dispersion of the resonance, as the former is directly associated with shapes of the hole and electron Fermi surfaces in reciprocal space \cite{JHZhang10}. 
From previous work on electron and hole-doped BaFe$_2$As$_2$, we know that the low-energy ($<40$ meV) normal state spin fluctuations change from transversely elongated for electron-doped BaFe$_{2-x}$Ni$_2$As$_2$ \cite{Luo13} to longitudinally elongated for hole doped Ba$_{0.67}$K$_{0.33}$Fe$_{2}$As$_{2}$, while the high-energy 
($E\ge 50$ meV) spin fluctuations of these materials have similar transverse elongation \cite{clzhang11,Meng13}. Therefore, it would be of great interests to
study the dispersion of the resonance in Ba$_{0.67}$K$_{0.33}$Fe$_{2}$As$_{2}$ and its electron-doping effect in  
Ba$_{0.67}$K$_{0.33}$(Fe$_{1-x}$Co$_{x}$)$_{2}$As$_{2}$ [Fig. 1(a)] to test the spin exciton hypothesis \cite{li12,goltz14}.

In this paper, we use time-of-flight (TOF) INS experiments to study wave vector and energy dependence of the resonance in  
hole-doped Ba$_{0.67}$K$_{0.33}$Fe$_{2}$As$_{2}$ ($T_c=38$ K) and its electron-compensated Ba$_{0.67}$K$_{0.33}$(Fe$_{0.92}$Co$_{0.08}$)$_{2}$As$_{2}$ ($T_c=28$ K) without static AF order, where Co and K doping levels are nominal [Fig. 1(a)]. We find 
that the resonance in Ba$_{0.67}$K$_{0.33}$Fe$_{2}$As$_{2}$ has 
a spin-wave ring-like dispersion extending more 
along the longitudinal ($H$) than the transverse ($K$) directions from ${\bf Q}_{\rm AF}$ [Fig. 2(a-c)]. Upon electron-doping to form Ba$_{0.67}$K$_{0.33}$(Fe$_{0.92}$Co$_{0.08}$)$_{2}$As$_{2}$ with reduced $T_c$, the dispersion along the longitudinal direction narrows [Figs. 1(c) and 1(d)]. These results can be understood as arising from isotropic superconducting gaps in nested hole and electron Fermi surfaces within the BCS theory in the random phase 
approximation (RPA) calculation of the spin exciton model \cite{maier08,Graser10}.
Our results thus establish that the spin-wave-like dispersion of the resonance
in iron pnictides is a 
spin exciton of a nested hole and electron Fermi surfaces.

\begin{figure}[t]
\includegraphics[scale=.3]{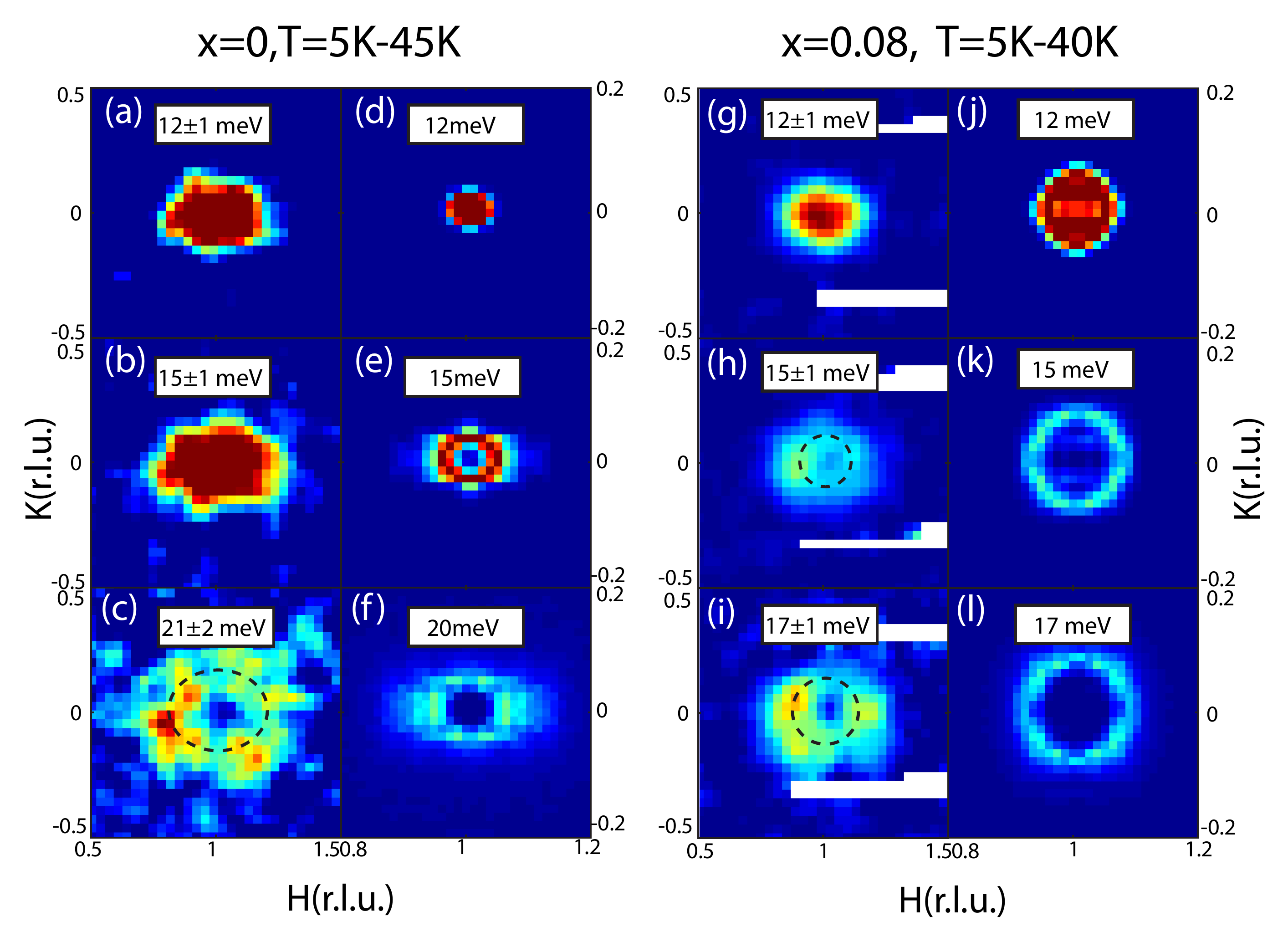}
\caption{Comparison of the dispersions of the resonance with BCS/RPA calculation.
(a-c) Constant energy slice of spin resonance for $x=0$. 
(d-f) Corresponding calculated images of the resonance from BCS/RPA theory.
(g-i) Constant energy slice of spin resonance For $x=0.08$.  The dashed curves in the figures
are expected spin wave dispersion using fits in Figs. 1(c), 1(d). 
(j-l) Corresponding images of the resonance from BCS/RPA theory.
 }
\end{figure}

We carried out INS experiments using the SEQUOIA spectrometer at the Spallation Neutron Source and the HB-3 triple-axis spectrometer in High Flux Isotope Reactor, both at Oak Ridge National Laboratory. For TOF INS experiment, we prepared 
11 g of sizable Ba$_{0.67}$K$_{0.33}$(Fe$_{0.92}$Co$_{0.08}$)$_{2}$As$_{2}$ single crystals and co-aligned them on aluminum plates \cite{clzhang11}. The Ba$_{0.67}$K$_{0.33}$Fe$_{2}$As$_{2}$ single 
crystals were previously measured \cite{Meng13}. The TOF experiments used incident neutrons parallel to the \textit{c} axis with incident energies of $E_i=35$, 80, and 250 meV with corresponding Fermi chopper 
frequency $\omega=180,420,600$ Hz, respectively. We define $(H,K,L)=(q_xa/2{\pi},q_yb/2{\pi},q_zc/2{\pi})$ using the orthorhombic lattice notation for the tetragonal lattice, 
where $a=b\approx 5.57$ \AA, $c=13.13$ \AA. The experiments
on HB-3 used a pyrolytic graphite monochromator, analyzer,
and filter after the sample with fixed final energy $E_f$=14.7meV and collimators of 
48$^\prime$–80$^\prime$-sample-40$^\prime$–240$^\prime$.

Figure 1(a) shows the electronic phase diagram of Ba$_{0.67}$K$_{0.33}$(Fe$_{1-x}$Co$_{x}$)$_{2}$As$_{2}$ as determined from our neutron diffraction experiments \cite{supplementary}. 
Consistent with earlier work \cite{li12,goltz14}, we see 
that Co-doping to Ba$_{0.67}$K$_{0.33}$Fe$_{2}$As$_{2}$ gradually suppresses 
superconductivity and induces long-range AF order for $x\geq 0.16$.
To systematically investigate the Co-doping evolution of the resonance without complication of
static AF order, we focus on $x=0, 0.08$ samples [Fig. 1(a)]. 
Figure 1(b) summarizes the three dimensional (3D) dispersion of the resonance in 
$x=0.08$, obtained by taking the temperature difference of 
spin excitation spectra between the superconducting state at $T=5$ K and the normal state at $T=40$ K. 
While the resonance first starts to emerge from $E=5$ meV at ${\bf Q}_{\rm AF}=(1,0)$, it has
strong in-plane dispersion along both the $H$ and $K$ directions, which leads to a ring of scattering in the $(H,K)$ plane at $E=18\pm 1$ meV. These results are clearly different from electron-doped 
Ba(Fe$_{0.963}$Ni$_{0.037}$)$_2$As$_2$ where the modes disperses only 
along the $K$ direction \cite{Kim13}. 

\begin{figure}[t]
\includegraphics[scale=.6]{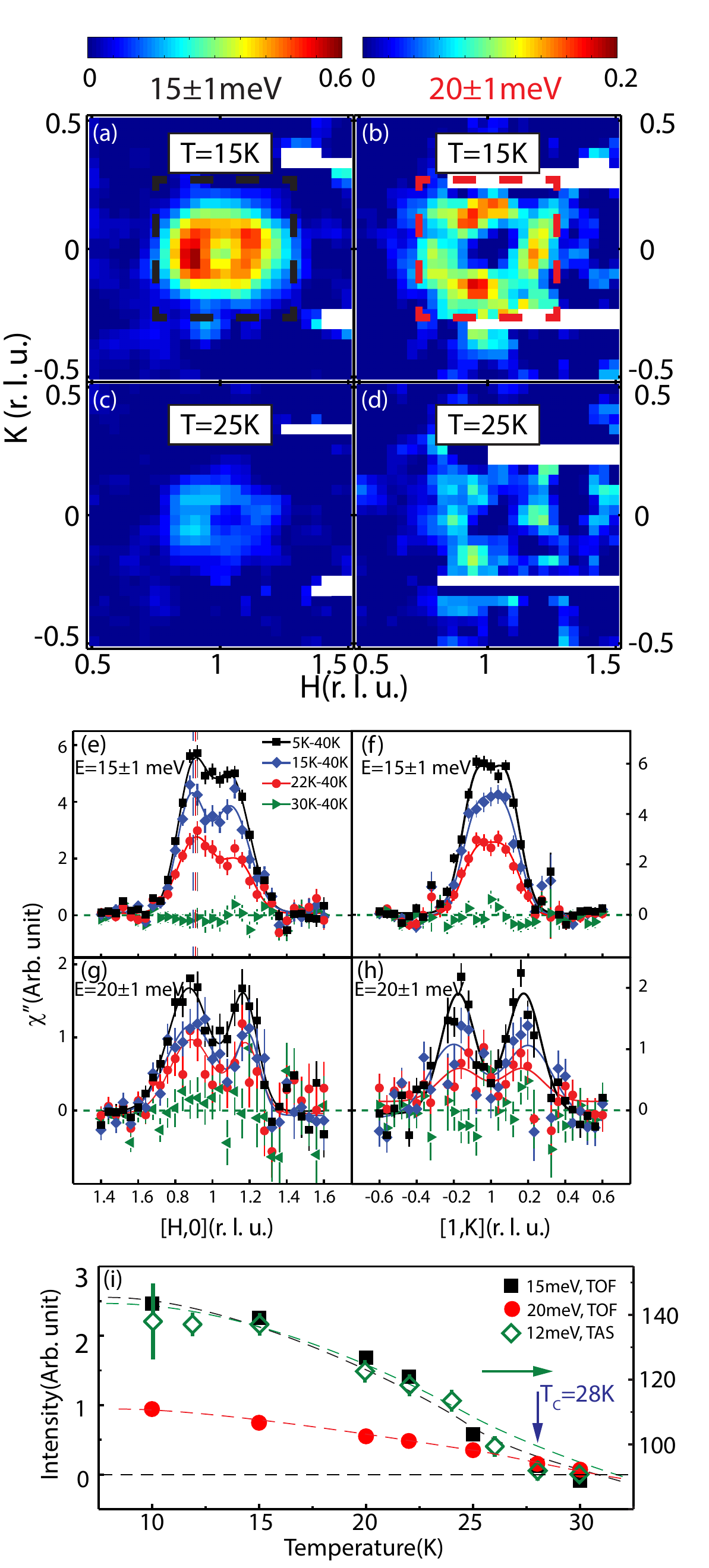}
\caption{Temperature dependence of the resonance at different energies for $x=0.08$. 
(a,c) Temperature dependence of the resonance at $E=15\pm 1$. (b,d) Identical scans
at $E=20\pm 1$ meV. (e-h) The corresponding 1D cuts along 
different directions. (i) Temperature dependence of the resonance at
different energies, where $T_c$ is marked by the vertical arrow. The integration area in $E=15\pm1$ and $20\pm1$ meV correspond to black and red dashed boxes in (a).
}
\end{figure}

To determine the Co-doping evolution of the resonance, we show in Figs. 1(c)
and 1(d) the dispersions of the mode along the $H$-direction for $x=0$ and 
0.08, respectively. 
Inspection of the figures reveals a clear narrowing of the width of the resonance with
increasing $x$. These results are qualitatively consistent with expectations of hole and electron
Fermi surface nesting, where Co-doping reduces the size of the hole pocket near $\Gamma$
and increases the size of the electron pocket near the $X/Y$ points [Figs. 1(e), and 1(f)] \cite{dai}.

Figure 2 summarizes the wave vector and energy dependence of spin excitations near the resonance energy in the normal
and superconducting states of $x=0,0.08$ samples.  For $x=0$, 
spin excitations in the superconducting  
state are longitudinally elongated, with the longitudinal elongation increasing  
with increasing energy [Figs. 2(a)-2(c)]. In the normal state [Figs. 2(d)-2(f)], the in-plane spin excitations show less anisotropy from 15 meV [Fig 2(d)] up to 23 meV [Fig 2(f)]. For $x=0.08$, while the normal state spin excitations are centered 
at ${\bf Q}_{\rm AF}=(1,0)$ for all measured 
energies [Figs. 2(j)-2(l)], progressive larger ring-like features 
appear at $E=13\pm 1, 17\pm 1,$ and $19\pm 1$ meV [Figs. 2(g)-2(i)].

To understand the normal state spin excitations and their connection with the resonance,
we fit the normal state spin excitations with a Fermi liquid model 
and find the in-plane two-dimensional (2D) correlation 
length $\xi_\textbf{q}$ \cite{Kim13,supplementary}. The anisotropic $\xi_\textbf{q}$ can be used to estimate the resonance dispersion $
\Omega_{\textbf{q}}^2=\Delta_{\textbf{q}}\Gamma_{\textbf{q}}(1+\xi_{\textbf{q}}^2{\bf q}^2)
$, where $\Delta_{\textbf{q}}$ is the $\textbf{q}$-dependent
superconducting gap and $\Gamma_{\textbf{q}}$ is 
the $\textbf{q}$-dependent Landau damping \cite{Kim13}.
Assuming an isotropic Landau damping and superconducting gap, the dispersion of the resonance mode is reduced to equation (1) with $c_{res,\textbf{q}}=\Omega_0\xi_\textbf{q}$, thus directly connects normal state spin correlation length (and its anisotropy) to the resonance dispersion. Although such a picture can qualitatively capture the ring-like upward dispersion of the resonance, it cannot explain the change in the longitudinal
elongation of the spin excitations from the 
normal to the superconducting state [Figs. 2(a)-2(f)] \cite{Meng13}.

By computing the differences between the normal ($T=45$ K) and superconducting 
($T=5$ K) state measurements \cite{eschrig}, Figures 3(a)-3(c) show in-plane 
$\textbf{q}$-dependence of
the resonance at energies of $E=12\pm 1, 15\pm 1,$ and $21\pm 2$ meV, respectively, for $x=0$. At $E=12\pm 1$ meV, the resonance is a longitudinally elongated ellipse centered at ${\bf Q}_{\rm AF}=(1,0)$. On moving to $E=15\pm 1$ meV, the ellipse becomes slightly larger but is still centered at ${\bf Q}_{\rm AF}$. Further increasing energies to 
$E=21\pm 2$, we find elliptical ring-like scattering dispersing 
away from ${\bf Q}_{\rm AF}$.  Figures 3(g)-3(i) summarize similarly 
subtracted data for $x=0.08$, which reveal a clear ring-like
resonance at energies $E=15\pm 1$, and $17\pm 1$ meV.  Compared 
with $x=0$, the resonance has ring-like scattering 
in $x=0.08$ but is more isotropic in reciprocal space 
along the $H$ and $K$ directions.

Although Figures 1-3 have shown the dispersive ring-like feature of the resonance in $x=0,0.08$ samples, 
the mode has a rather broad energy width that may be related to inhomogeneous superconductivity \cite{eschrig}. It is therefore important to establish temperature dependence of the commensurate and
ring-like response below $T_c$, and determine if the ring-like feature
also responds to superconductivity and is related to the superconducting gap function.
Figures 4(a)-4(d) show the 2D images of the resonance at 15 K and  25 K
for the $x=0.08$ sample. The corresponding one-dimensional (1D) cuts are shown in Figs. 4(e)-4(h). While
intensity of the resonance at probed energies decreases with increasing temperature and
vanishes at $T_c$, the wave vector dependence of the mode and the ring-like feature have no
visible temperature dependence. 
Figure 4(i) shows temperature dependence of the integrated
intensity of the resonance at $E=15\pm 1, 20\pm 1,$ and 12 meV. At all probed energies, temperature
dependence of the resonance behave identically, suggesting that they are related to the superconducting gap function.

Having established the wave vector and Co-doping dependence of the resonance dispersion and normal
state spin excitations in 
Ba$_{0.67}$K$_{0.33}$(Fe$_{1-x}$Co$_{x}$)$_{2}$As$_{2}$, we now test if the dispersion of the mode is well described by the spin exciton model of eq. (1) \cite{Kim13}. To do this, we first fitted the 2D normal state spin excitations in Figs. 2(d)-2(f) and 2(j)-2(l) with 
 $\xi_\textbf{q}$ \cite{Kim13,supplementary}, the outcome was   
then used to fit the data in the superconducting state and obtain $c_{res,\textbf{q}}$ along different directions. For $x=0$, we find $c_{res,H}\approx 154$ and $c_{res,K}\approx 169$ meV\AA.
Fitting the actual dispersion curves of the resonance in Figs. 1(c) and 3(a)-3(c) 
with a linear dispersion yields $c_{res,H}({\rm exp}) \approx 65$ and $c_{res,K}({\rm exp})\approx 84$ meV\AA. 
Similarly, we find $c_{res,H}\approx 126$ and $c_{res,K}\approx 141$ meV\AA, and 
$c_{res,H}({\rm exp}) \approx 78$ and $c_{res,K}({\rm exp})\approx 87$ meV\AA\ for $x=0.08$. The effect of increasing Co-doping from $x=0$ to $x=0.08$
is to increase $c_{res,H}({\rm exp})$, while $c_{res,K}({\rm exp})$ remains virtually unchanged.

To quantitatively understand the experimental results, we have used a BCS/RPA approximation \cite{maier08} to calculate the magnetic susceptibility $\chi^{\prime\prime}({\bf Q}_{\rm AF},E)$ from a 3D tight-binding 
five-orbital Hubbard-Hund model that describes the electronic structure of BaFe$_2$As$_2$ \cite{Graser10}. The effect of doping by K and Co substitution is estimated by a rigid band shift. Specifically, we use a filling of $\langle n\rangle = 5.915$ corresponding to a hole doping of 8.5\% to model the $x=0.08$ system. For the superconducting gap, we have used an isotropic $s^\pm$ gap with $\Delta = 8\, {\rm meV}$ on the Fermi surface hole cylinders around the zone center and $\Delta = -8\, {\rm meV}$ on the electron cylinders around the zone corner. The interaction matrix in orbital space used in the RPA calculation contains on-site matrix elements for the intra-orbital and inter-orbital Coulomb 
repulsions $U$ and $U^\prime$, and for the Hunds-rule coupling and pair-hopping terms $J$ and $J^\prime$, respectively. Here, we have used spin-rotationally invariant 
parameters $J = J^\prime = U/4$ and $U^\prime = U/2$ with $U = 0.77\, {\rm eV}$.

For these parameters, we obtain a resonance in $\chi^{\prime\prime}({\bf Q}_{\rm AF},E)$ at ${\bf Q}_{\rm AF}=(1,0)$ and $E = 10\, {\rm meV}$. Moving away from ${\bf Q}_{\rm AF}$, the resonance disperses upward resulting in a ring-like feature in constant energy scans that is slightly elongated along the longitudinal direction similar to what is observed in the experimental data. At energies above $\sim 17\, {\rm meV}$, the ring-like excitations disappear and change into a broad 
blob centered at ${\bf Q}_{\rm AF}=(1,0)$. Figures 3(d)-3(f) and 3(j)-3(l) summarize the 
Co-doping and energy dependence of the resonance 
from the RPA calculation. we see that the RPA calculation with isotrpoic 
superconducting 
gap can describe very well the energy and doping evolution of the resonance, further 
confirming the spin exciton nature of the resonance although details of the dispersion calculated
from RPA still differ somewhat from the experiments. 

In summary, we have used TOF INS to study the wave vector-energy dispersion
of the resonance in Ba$_{0.67}$K$_{0.33}$(Fe$_{1-x}$Co$_{x}$)$_{2}$As$_{2}$ with $x=0,0.08$. Compared with electron-doped underdoped superconducting Ba(Fe$_{0.963}$Ni$_{0.037}$)$_2$As$_2$, where the resonance displays a strong transverse dispersion but centered at ${\bf Q}_{\rm AF}$ along the longitudinal direction \cite{Kim13}, the resonance in Ba$_{0.67}$K$_{0.33}$(Fe$_{1-x}$Co$_{x}$)$_{2}$As$_{2}$ has ring-like dispersion that follows the evolution of the Fermi surface nesting with increasing
Co-doping. These results are consistent with expectations of a spin exciton model with BCS/RPA approximation, indicating that the mode arises from 
particle-hole excitations involving momentum states near the sign-reversed electron-hole Fermi surfaces.

The neutron-scattering work at Rice University was supported by the US NSF Grant
No. DMR-1700081 (P.D.). The single-crystal
synthesis work was supported by the Robert
A. Welch Foundation Grant No. C-1839 (P.D.). T.A.M was supported by the U.S. Department of Energy, Office of Basic Energy Sciences, Materials Sciences and Engineering Division. This research used resources at the High Flux Isotope Reactor and Spallation Neutron Source, a DOE Office of Science User Facility operated by the Oak Ridge National Laboratory.


\begin{thebibliography}{}

\bibitem{scalapino} D. J. Scalapino, Rev. Mod. Phys. {\bf 84}, 1383 (2012).

\bibitem{BKeimer} B. Keimer, S. A. Kivelson, M. R. Norman, S. Uchida, J. Zaanen, Nature (London) {\bf 518}, 179-186 (2015).

\bibitem{dai} P. C. Dai, Rev. Mod. Phys. {\bf 87}, 855 (2015).

\bibitem{thompson} J. D. Thompson and Z. Fisk, J. Phys. Soc. Jpn. {\bf 81}, 011002 (2012).

\bibitem{J.Rossat-Mignod} J. Rossat-Mignod, L. P. Regnault, C. Vettier, P. Bourges, P. Burlet, J. Bossy, J. Y. Henry, and G. Lapertot, Phys. C \textbf{185-189}, 86 (1991). 

\bibitem{eschrig} M. Eschrig, Adv. Phys. {\bf 55}, 47–183 (2006).

\bibitem{Abanov99} A. Abanov and A. V. Chubukov  Phys. Rev. Lett. {\bf 83}, 1652 (1999). 

\bibitem{chubukov01} A.V. Chubukov, B. Janko, and O. Tchernyshyov, Phys. Rev. B {\bf 63}, 180507(2001).

\bibitem{christianson}
A. D. Christianson, E. A. Goremychkin, R. Osborn, S. Rosenkranz, M. D. Lumsden, C. D. Malliakas, I. S. Todorov, H. Claus, D. Y. Chung, M. G. Kanatzidis, R. I. Bewley, and T. Guidi,
Nature {\bf 456}, 930 (2008).

\bibitem{lumsden} M. D. Lumsden, A. D. Christianson, D. Parshall, M. B. Stone, S. E. Nagler, G. J. MacDougall, H. A. Mook, K. Lokshin, T. Egami, D. L. Abernathy, E. A. Goremychkin, R. Osborn, M. A. McGuire, A. S. Sefat, R. Jin, B. C. Sales, and D. Mandrus,
Phys. Rev. Lett. {\bf 102}, 107005 (2009).

\bibitem{schi09} S. Chi, A. Schneidewind, J. Zhao, L. W. Harriger, L. J. Li, Y. K. Luo, G. H. Cao, Z. A. Xu, M. Loewenhaupt, J. P. Hu, and P. C. Dai,
Phys. Rev. Lett. {\bf 102}, 107006 (2009).

\bibitem{dsinosov09}	
D. S. Inosov, J. T. Park, P. Bourges, D. L. Sun, Y. Sidis, A. Schneidewind, K. Hradil, D. Haug, C. T. Lin, B. Keimer, and V. Hinkov,
  Nat. Phys. {\bf 6}, 178 (2010).

\bibitem{C.Zhang} C. Zhang, R. Yu, Y. Su, Y. Song, M. Wang, G. Tan, T. Egami, J. A. Fernandez-Beca, E. Faulhaber, Q. Si, and P. C. Dai, Phys. Rev. Lett. \textbf{111}, 207002 (2013).

\bibitem{stock08} C. Stock, C. Broholm, J. Hudis, H. J. Kang, and C. Petrovic, Phys. Rev. Lett. {\bf 100}, 087001 (2008). 

\bibitem{wilson} S. D. Wilson, P. C. Dai, S. L. Li, S. X. Chi, H. J. Kang, and J. W. Lynn, Nature {\bf 442}, 59 (2006). 

\bibitem{D.Inosov} D. S. Inosov, J. T. Park, A. Charnukha, Y. Li, A. V. Boris, B. Keimer, and V. Hinkov, Phys. Rev. B \textbf{83}, 214520 (2011).

\bibitem{G.Yu} G. Yu, Y. Li, E. M. Motoyama, and M. Greven, Nat. Phys. \textbf{5}, 873 (2009).

\bibitem{QSWang16} Qisi Wang, J. T. Park, Yu Feng, Yao Shen, Yiqing Hao, Bingying Pan, J.W. Lynn, A. Ivanov, Songxue Chi, M. Matsuda, Huibo Cao, R. J. Birgeneau, D. V. Efremov, and Jun Zhao, Phys. Rev. Lett. {\bf 116}, 197004 (2016).

\bibitem{bourges00} P. Bourges, Y. Sidis, H. F. Fong, L. P. Regnault, J. Bossy, A. Ivanov, B. Keimer, 
Science {\bf 288}, 1234 (2000). 

\bibitem{dai00} P. C. Dai, H. A. Mook, R. D. Hunt, and F. Do$\rm \breve{g}$an, Phys. Rev. B {\bf 63}, 054525 (2001).

\bibitem{Reznik04} D. Reznik, P. Bourges, L. Pintschovius, Y. Endoh, Y. Sidis, T. Masui, and S. Tajima, Phys. Rev. Lett. {\bf 93}, 207003 (2004).

\bibitem{Hayden04} S. M. Hayden, H. A. Mook, P. C. Dai, T. G. Perring, and F. Do$\rm \breve{g}$an, Nature {\bf 429}, 531 (2004).

\bibitem{xylu13} X. Lu, H. Gretarsson, R. Zhang, X. Liu, H. Luo, W. Tian, M. Laver, Z. Yamani, Y. -J. Kim, A. H. Nevidomskyy, Q. Si, and P. Dai, Phys. Rev. Lett. {\bf 110}, 257001 (2013).

\bibitem{Kim13} M. G. Kim, G. S. Tucker, D. K. Pratt, S. Ran, A. Thaler, A. D. Christianson, K. Marty, S. Calder,
A. Podlesnyak, S. L. Bud'ko, P. C. Canfield, A. Kreyssig, A. I. Goldman, and R. J. McQueeney, Phys. Rev. Lett. {\bf 110}, 177002 (2013).

\bibitem{raymond15} S. Raymond and G. Lapertot, Phys. Rev. Lett. {\bf 115}, 037001 (2015).

\bibitem{song16} Yu Song, John Van Dyke, I. K. Lum, B. D. White, Sooyoung Jang, Duygu Yazici, L. Shu,
A. Schneidewind, Petr $\rm \check{C}$erm$\rm \acute{a}$k, Y. Qiu, M. B. Maple, Dirk K. Morr, and P. C. Dai,
Nat. Comm. {\bf 7}, 12774 (2016).

\bibitem{Das14} Pinaki Das, S.-Z. Lin, N. J. Ghimire, K. Huang, F. Ronning, E. D. Bauer, J. D. Thompson, C. D. Batista, 
G. Ehlers, and M. Janoschek, Phys. Rev. Lett. {\bf 113}, 246403 (2014).

\bibitem{stock15} C. Stock, J. A. Rodriguez-Rivera, K. Schmalzl, E. E. Rodriguez, A. Stunault, and C. Petrovic, 
Phys. Rev. Lett. {\bf 114}, 247005 (2015).

\bibitem{hirschfeld} P. J. Hirschfeld, M. M. Korshunov, and I. I. Mazin, Rep. Prog. Phys. {\bf 74}, 124508 (2011).

\bibitem{chubukov08} A. V. Chubukov and L. P. Gor'kov, Phys. Rev. Lett. {\bf 101}, 147004 (2008).

\bibitem{tucker12} G. S. Tucker, R.M. Fernandes, H.-F. Li, V. Thampy, N. Ni, D. L. Abernathy, S. L. Bud’ko, P. C. Canfield, D. Vaknin, J. Schmalian, and R. J. McQueeney, Phys. Rev. B {\bf 86}, 024505 (2012).

\bibitem{Luo12} H. Q. Luo, Z. Yamani, Y. C. Chen, X. Y. Lu, M. Wang, S. L. Li, T. A. Maier, S. Danilkin, D. T. Adroja and P. C. Dai, Phys. Rev. B {\bf 66}, 024508(2012).

\bibitem{JHZhang10} J. H. Zhang, R. Sknepnek, and J. Schmalian, 
Phys. Rev. B {\bf 82}, 134527 (2010).

\bibitem{Luo13} Huiqian Luo, Xingye Lu, Rui Zhang, Meng Wang, E. A. Goremychkin, D. T. Adroja, Sergey Danilkin, Guochu Deng, Zahra Yamani, and Pengcheng Dai, Phys. Rev. B {\bf 88}, 144516 (2013).

\bibitem{clzhang11} C. L. Zhang, M. Wang, H. Q. Luo, M. Y. Wang, M. S. Liu, J. Zhao, D. L. Abernathy, T. A. Maier, Karol Marty, M. D. Lumsden, S. Chi, S. Chang,
Jose A. Rodriguez-Rivera, J. W. Lynn, T. Xiang, J. P. Hu, and P. C. Dai, 
Sci. Rep. {\bf 1}, 115 (2011).

\bibitem{Meng13} Meng Wang, Chenglin Zhang, Xingye Lu, Guotai Tan, Huiqian Luo, Yu Song, Miaoyin Wang, Xiaotian Zhang, E.A. Goremychkin, T.G. Perring, T.A. Maier, Zhiping Yin, Kristjan Haule, Gabriel Kotliar,	and Pengcheng Dai, Nat. Comm. {\bf 4}, 2874 (2013).

\bibitem{li12} J. Li, Y. F. Guo, S. B. Zhang, J. Yuan, Y. Tsujimoto, X. Wang, C. I. Sathish, Y. Sun, S. Yu, W. Yi, K. Yamaura, E. Takayama-Muromachiu, Y. Shirako, M. Akaogi, and H. Kontani, Phys. Rev. B {\bf 85}, 214509 (2012) 

\bibitem{goltz14} T. Goltz, V. Zinth, D. Johrendt, H. Rosner, G. Pascua, H. Luetkens, P. Materne, and H-H. Klauss, Phys. Rev. B {\bf 89}, 144511 (2014).

\bibitem{maier08} T. Maier and D. Scalapino, Phys. Rev. B {\bf 78}, 020514 (2008).

\bibitem{Graser10} S. Graser, A. F. Kemper, T. A. Maier, H. P. Cheng, P. J. Hirschfeld, and D. J. Scalapino,  Phys. Rev. B {\bf 81}, 214503 (2010). 

\bibitem{supplementary}  For detailed data analysis and additional transport results, see supplementary material.
 


\end{thebibliography}
\end{document}